# Far-field Super-resolution Imaging with Dual-Dye-Doped Nanoparticles

Jianfang Chen [1, †] and Ya Cheng [2]

[1] *Shanghai Institute of Optics and Fine Mechanics, Chinese Academy of Sciences*
*P.O. Box 800-211, Shanghai 201800, China*

[2] *State Key Laboratory of High Field Laser Physics, Shanghai Institute of Optics and Fine Mechanics, Chinese Academy of Sciences*
*P.O. Box 800-211, Shanghai 201800, China*

[†] Corresponding author: cjianfang@hotmail.com

**Abstract:**

We propose to achieve super-resolution in far-field imaging by use of dual-dye-doped nanoparticles. The nanoparticles with a diameter of a few nanometers are co-doped with two types of dye molecules, namely, Cy3 and Cy5, at a controllable ratio. Due to the short distances between the dye molecules confined in the nanoparticles, Förster resonant energy transfer can occur between the Cy3 and Cy5 molecules with high efficiency. Therefore, the Cy5 molecules can quench the fluorescence emission from the Cy3 molecules in the outer region of focal spot of the excitation beam, thereby enhancing the resolution of imaging.

***OCIS codes***: 180.0180, 100.6640, 100.6890.

In recent years, three-dimensional (3D) far-field optical imaging with a resolution beyond diffraction limit has attracted tremendous attention due to its potential for achieving electron microscope resolution in 3D living cell imaging [1]. To this end, many different approaches have been proposed, such as stimulated emission depletion microscopy (STED) [2], stochastic optical reconstruction microscopy (STORM) [3], photoactivatable localization microscopy (PALM) [4], saturated pattern excitation microscopy (SPEM) [5], 4Pi microscopies [6, 7], and so on. Currently, many super-resolution far-field imaging techniques have been successfully demonstrated [2-6], while it has been pointed out that further efforts are still required in order to overcome several important issues, such as limited imaging speed, photobleaching of dye, low fluorescence flux, and so forth [1]. In addition, one may also have noticed that the current far-field super-resolution microscopes are generally more complex and more expensive than the traditional scanning fluorescence microscope in terms of focal geometry [5-7] and excitation light source [2-4]. In this Letter, we theoretically show that super-resolution in 3D far-field imaging can be achieved with a traditional scanning fluorescence microscope and merely using linear fluorescence excitation. The trick to beating the diffraction limit is to employ specially designed dual-dye-doped nanoparticles as fluorescent labels, which show an extremely nonlinear optical behavior under the light illumination.

To begin with, we first illustrate the dual-dye-doped nanoparticle in Fig. 1. We assume that the nanoparticle is a nanoscale silica sphere in which two kinds of dyes are co-doped between which Förster resonant energy transfer (FRET) can occur. Particularly, to facilitate a numerical analysis, we assume that Cy3 and Cy5 dye molecules are used for constructing the dual-dye-doped nanoparticles, although in principle other FRET pairs can also be chosen. According to FRET theory, the rate of energy transfer is given by the equation,

$$k_F = (1/\tau_D)\cdot[R_0/r]^6, \quad [1]$$

where $R_0$ is the Förster critical distance, defined as the acceptor-donor separation radius for which the transfer rate equals the rate of donor decay in the absence of acceptor. Specifically, for the Cy3-Cy5 dye pair, it is know that $R_0 > 5nm$. In addition, $\tau_D$ is the donor lifetime in the absence of the acceptor, and $r$ is the distance separating the donor and acceptor. From Eq. (1), it is obvious that if the distance between the Cy3 and Cy5 dye molecules is sufficiently short as compared to $R_0$, FRET process can occur very efficiently. To fulfill this condition, we assume that the diameter of the silica nanoparticle in Fig. (1) is less than $R_0/2$. Thus, the distance between any two dye molecules encapsulated in the silica nanoparticle can be determined as $r \le R_0/2$. In this case, the rate of energy transfer from any donor molecule to an acceptor in the silica nanoparticle can be determined by Eq. (1), which is $k_F > (64/\tau_D) >> 1/\tau_D$. It is noteworthy that recently, FRET silica nanoparticles co-coped with more than one type of dye molecules have been synthesized and multi-color emission under single wavelength excitation has been experimentally demonstrated [9, 10]. By use of the similar approach, the silica nanoparticles as shown in Fig. (1) should be able to be produced.

We now examine the optical property of the dual-dye-doped nanoparticle under the linear fluorescence excitation. We define the spontaneous emission lifetimes of donor (Cy3) and acceptor (Cy5) as $\tau_D$ and $\tau_A$, respectively. In particular, the excitation wavelength is carefully selected so as to excite only the donor molecule while the excitation of the acceptor molecule can be efficiently avoided, owing to the separated excitation spectral peaks between the donor and acceptor molecules. The laser intensity used for excitation of the donor is defined as $I_E$. The excitation cross section of the donor is defined as $\sigma_E$. Figure 2 shows that under these conditions, the Cy3 dye can first be pumped to the excited state $S_1$, and then rapidly decay to the lowest vibrational energy level of the excited state through vibrational relaxation in a picosecond timescale. After this, the Cy3 dye molecule can return to its ground state $S_0$ either by resonantly transferring its energy to the Cy5 molecule or by spontaneously emitting a fluorescent photon. It is noteworthy that the FRET process can only occur when some Cy5 molecules still remain in the ground state; otherwise the FRET process will stop due to the depleted acceptor molecules in the ground state. The dynamics of fluorescence excitation and energy transfer between the donor and acceptor is illustrated in Fig. (2). Accordingly, the rate equation can be written as

$$\begin{cases} \dfrac{dn_D^E}{dt} = n_D^G I_E \sigma_E - n_D^E (k_D + k_F) \\ \dfrac{dn_A^E}{dt} = n_D^E k_F - n_A^E k_A \end{cases}, \quad if \quad n_A^G \geq n_D^E, \tag{2}$$

$$\begin{cases} \dfrac{dn_D^E}{dt} = n_D^G I_E \sigma_E - n_D^E k_D - n_A^G k_F \\ \dfrac{dn_A^E}{dt} = n_A^G k_F - n_A^E k_A \end{cases}, \quad if \quad n_A^G \leq n_D^E, \tag{3}$$

$$n_D^G + n_D^E = N_D, \tag{4}$$

$$n_A^G + n_A^E = N_A, \tag{5}$$

where $n_{D(A)}^E$ and $n_{D(A)}^G$ are the number of donor (acceptor) molecules in the excited and ground states, respectively; and $N_{D(A)}$ is the total number of donor (acceptor) molecules. In addition, $k_D = 1/\tau_D$, $k_A = 1/\tau_A$, and $k_F$ are the rates of spontaneous emission from donor, acceptor, and the rate of energy transfer from donor to acceptor, respectively. It is noteworthy that in deriving Eqs. [2-5], we have assumed that the vibrational decay rates for both the donor and acceptor are much higher than the $k_D$, $k_A$, and $k_F$; namely, the donor and acceptor molecules are assumed to relax to the lowest vibrational energy levels of their excited states almost instantly after being excited. This condition is generally fulfilled for many dyes as the typical time scale of the vibrational decay is on the picosecond level, whereas the time scale of the spontaneous fluorescence emission is typically on the nanosecond level. For continuous wave (CW) excitation, the fluorescence excitation will finally reach a steady state, thus the time derivatives in Eqs. [2-3] are zero. The analytical solution of Eqs. [2-5] can then be found as

$$\begin{cases} n_D^E = N_D I_E \sigma_E / (I_E \sigma_E + k_D + k_F), \quad if \quad n_A^G = N_A - (k_F / k_A) n_D^E \geq n_D^E \\ n_D^E = [N_D I_E \sigma_E - (N_A k_A k_F)/(k_A + k_F)]/(I_E \sigma_E + k_D), \quad if \quad n_A^G = N_A k_A / (k_A + k_F) \leq n_D^E \end{cases} \tag{6}$$

As have been mentioned above, due to the high efficiency of the energy transfer between the donor and acceptor molecules confined in the tiny nanoparticles, we have $k_F > (64/\tau_D) >> k_D$. Therefore, Eq. (6) can be further simplified to

$$\begin{cases} n_D^E \approx N_D I_E \sigma_E / k_F, \quad if \quad I_E \leq (k_A N_A)/(\sigma_E N_D) \\ n_D^E \approx [N_D I_E \sigma_E - N_A k_A]/(I_E \sigma_E + k_D), \quad if \quad I_E \geq (k_A N_A)/(\sigma_E N_D) \end{cases} \tag{7}$$

Since the intensity of fluorescence emitted from the donor molecules is determined by $I_F \propto n_D^E$, Equation (7) clearly shows that at a relatively low excitation intensity [i.e., $I_E \leq (k_A N_A)/(\sigma_E N_D)$], the fluorescence from the donor is very weak because of the facts that $k_F >> k_A$ (we assume that the fluorescence from acceptor is filtered). However, when the excitation intensity reaches $I_E = (k_A N_A)/(\sigma_E N_D)$, the donor fluorescence starts to grow sharply. The normalized fluorescence intensity as a function of excitation intensity is plotted in Fig. 2(b) by use of Eqs. [6]. In the calculation, the parameters are assumed as $N_A / N_D = 1$ and $k_F = 64k_D = 211k_A$ . The ratio between $k_D$ and $k_A$ is determined from their fluorescence lifetimes (for Cy3 and Cy5 molecules are ~0.3 ns and ~1 ns, respectively). The solid curve in Fig. 2(b) clearly shows that when the excitation intensity increases to near $I_E = (k_A N_A)/(\sigma_E N_D)$, the fluorescence from the donor molecules responds to the excitation light in an extremely nonlinear manner. It is this extraordinary optical property of the dual-dye-doped silica nanoparticle which can enable the super-resolution in 3D far-field imaging, as we can see below.

In Figs. 3(a) and 3(b), we compare the point-spread functions (PSFs) of a conventional scanning fluorescence microscope equipped with an oil immersion objective with a numerical aperture (NA) of 1.40 (refractive index of oil: 1.51; semi-aperture angle: 67°) for two different kinds of fluorescent labels, namely, the Cy3 molecules and the dual-dye-doped silica nanoparticles, respectively. For both cases, we choose an excitation wavelength of 510 nm at which a high excitation efficiency of Cy3 molecules can be warranted whereas the excitation of Cy5 molecules can essentially be avoided (thus we can neglect the excitation of acceptor). Furthermore, the peak intensity at the center of the focal spot of the excitation beam is chosen to be $I_P = 0.5k_D / \sigma_E$. Under these conditions, Fig. 3(a) shows the familiar PSF of the conventional scanning fluorescence microscope when the Cy3 dyes is used as the fluorescent labels; whereas in Fig. 3(b), the PSF of the same scanning fluorescence microscope, which instead uses the dual-dye-doped nanoparticles as labels in this case, is significantly size-reduced, due to the nonlinear optical behavior described by Eq. (6). The lateral and longitudinal resolutions (full with at half maximum, FWHM) for the PSF shown in Fig. 3(b) are ~112 nm and ~306 nm, respectively, which are significantly improved in comparison with the lateral and longitudinal resolutions of ~213 nm and ~580 nm of the PSF shown in Fig. 3(a), respectively.

In summary, we have shown that the use of dual-dye-doped nanoparticles as fluorescent labels can allow for achieving super-resolution in a conventional scanning fluorescence microscope. The system is easy to operate and cost-effective. The basic concept is to employ a controlled FRET process in which the acceptor dye molecules are used to quench the fluorescence from the donor molecules when the excitation intensity is below a threshold. In principle, the imaging resolution which could be achievable with this technique is only limited by the size of the nanoparticles; however, in practice, as one may have already noticed that in order to improve the resolution to its maximum, the excitation intensity must be chosen as close to $I_E = (k_A N_A)/(\sigma_E N_D)$ as possible, which in turn results in a significantly weakened fluorescence emission. Therefore, the ultimate limitation on the achievable resolution with this technique is the signal to noise ratio. Nevertheless, our simulation has shown that moderately improved spatial resolution can still be obtained with an acceptable fluorescence signal loss. Additionally, one should note that the bandwidth of Fourier frequency spectrum of the PSF in Fig. 3(b) is significantly enlarged, implying that the spatial resolution can be further improved by Fourier deconvolution. Finally, we point out that since the dual-dye-doped nanoparticle responses weakly to a low-intensity excitation, a single-photon-excitation based 3D scanning fluorescence microscope can thus be constructed without a confocal pinhole because of the depleted fluorescence emission in the out-of-focus areas.

**References:**

[1] S. W. Hell, "Far-Field Optical Nanoscopy", Science **316**, 1153 (2007).

[2] M. Dyba and S. W. Hell, "Focal Spots of Size λ/23 Open Up Far-Field Florescence Microscopy at 33 nm Axial Resolution", Phys. Rev. Lett. **88**, 163901 (2002)

[3] E. Betzig, G. H. Patterson, R. Sougrat, O. W. Lindwasser, S. Olenych, J. S. Bonifacino, M. W. Davidson, J. Lippincott-Schwartz, H. F. Hess, "Imaging Intracellular Fluorescent Proteins at Nanometer Resolution", Science **313**, 1462 (2006).

[4] M. J. Rust, M. Bates and X. W. Zhuang, "Sub-diffraction-limit imaging by stochastic optical reconstruction microscopy (STORM)", Nat. Methods **3**, 793 (2006)

[5] M. G. L. Gustafsson, "Nonlinear structured-illumination microscopy: Wide-field fluorescence imaging with theoretically unlimited resolution", Proc. Natl. Acad. Sci. U.S.A. **102**, 13081 (2005).

[6] M. C. Lang, J. Engelhardt, and S. W. Hell, "4Pi microscopy with linear fluorescence excitation," Opt. Lett. **32**, 259 (2007)

[7] J. Chen, and K. Midorikawa, "Two-color Two-photon 4Pi microscopy", Opt. Lett. **29**, 1354 (2004)

[8] Z. Jiang, and W. A. Goedel, "Fluorescence properties of systems with multiple Förster transfer pairs", Phys. Chem. Chem. Phys. **10**, 4584 (2008)

[9] L. Wang, C. Yang,and W. Tan, "Dual luminophore-doped silica nanoparticles for multiplexed signaling", Nano Lett. **5**, 37 (2005)

[10] L. Wang, and W. Tan, "Multicolor FRET silica nanoparticles by single wavelength excitation", Nano Lett. **6,** 84 (2006)

**Captions of Figures:**

Fig. 1. (Color online) Illustration of a dual-dye-doped silica nanoparticle.

Fig. 2. (Color online) (a) Schematic of dynamics showing fluorescence excitation of donor and resonant energy transfer between donor and acceptor. (b) Normalized fluorescence intensities as functions of the excitation intensity. Solid curve: Cy3 molecule in the dual-dye-doped nanoparticle as fluorescent label; dashed curve: Cy3 molecule as fluorescent label.

Fig. 3. (Color online) PSFs of a conventional 3D scanning fluorescence microscope using (a) Cy3 dye molecules and (b) dual-dye-doped nanoparticles as fluorescent labels.

Fig. 1

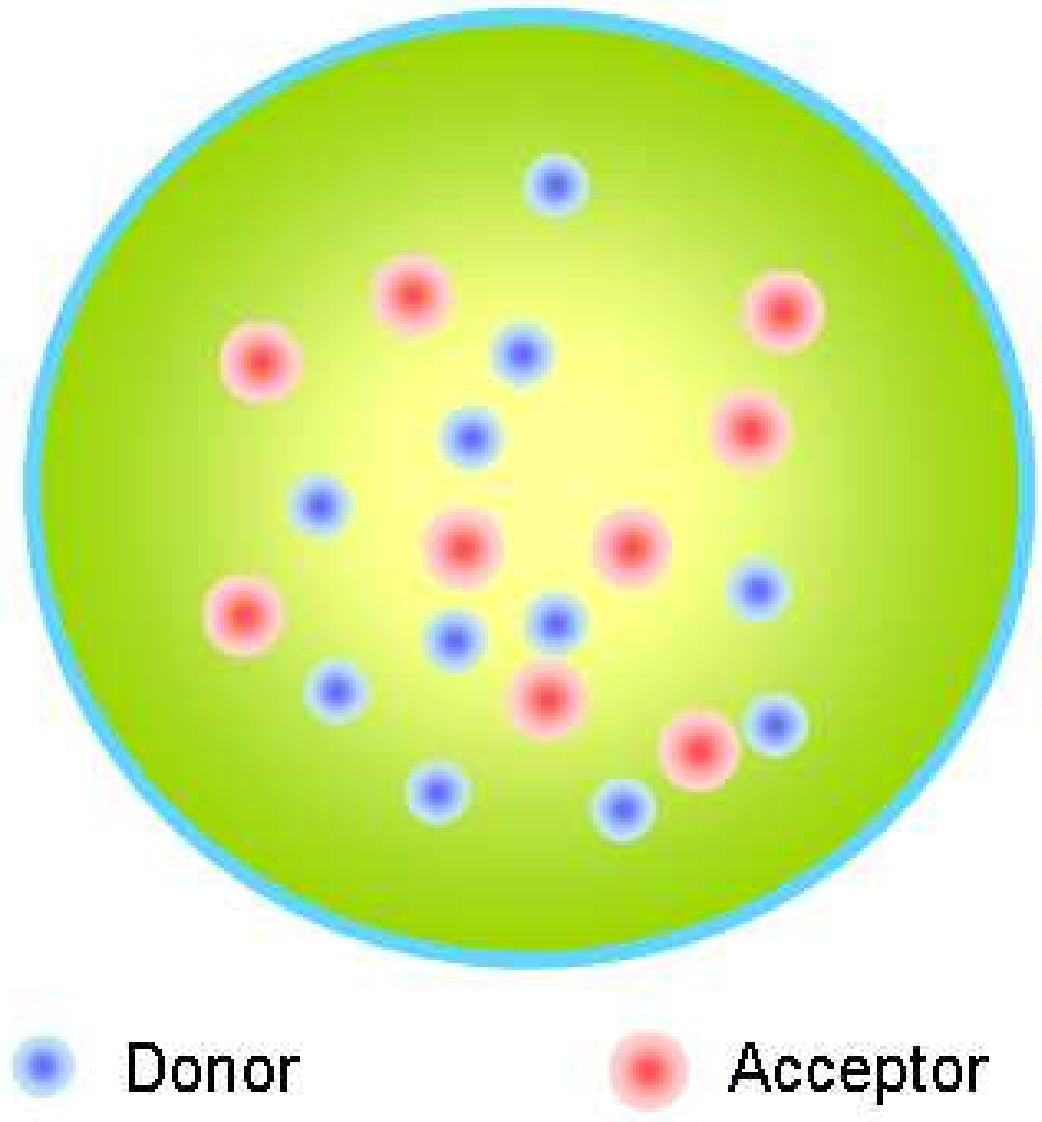

Fig. 2

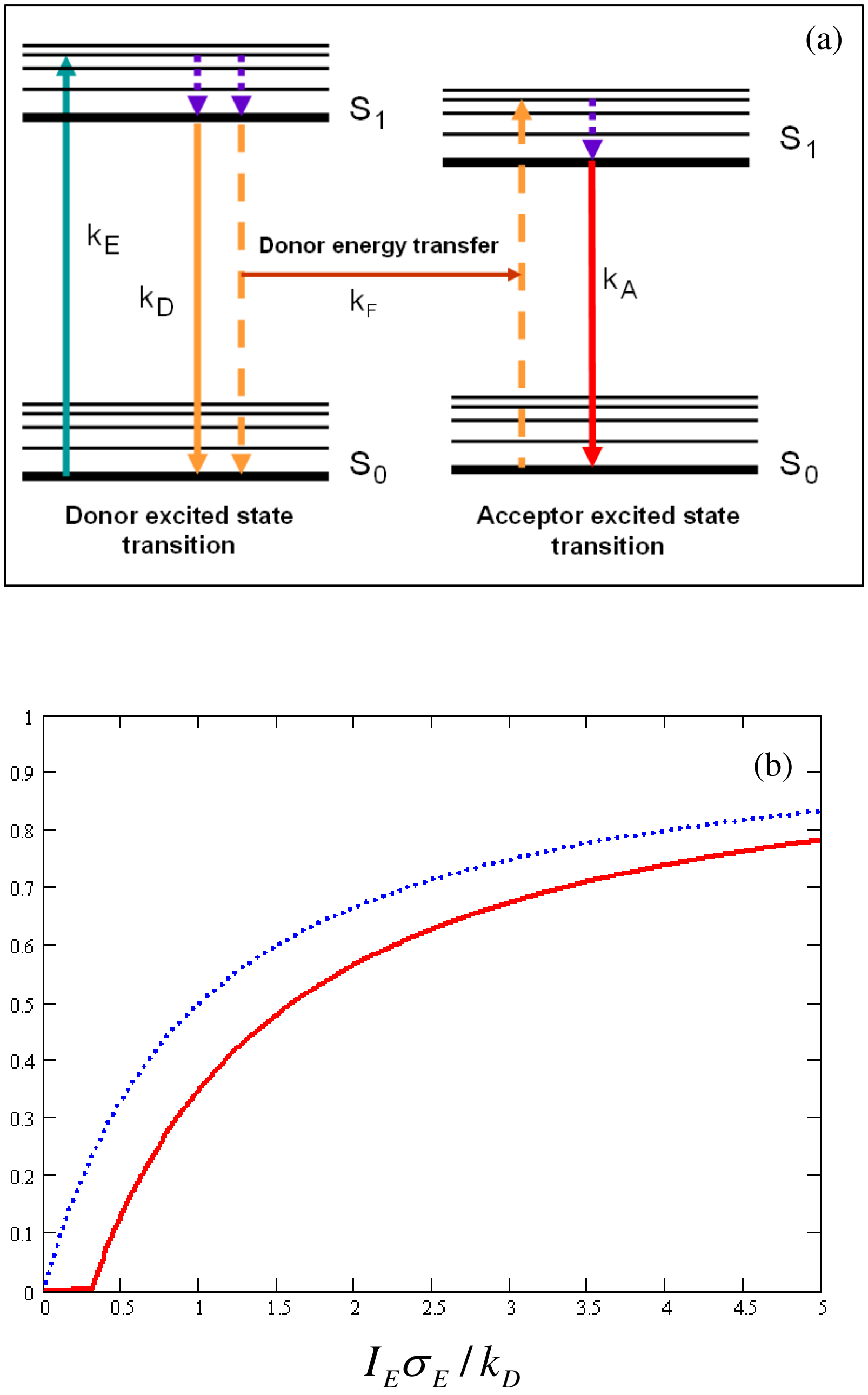

Fig. 3

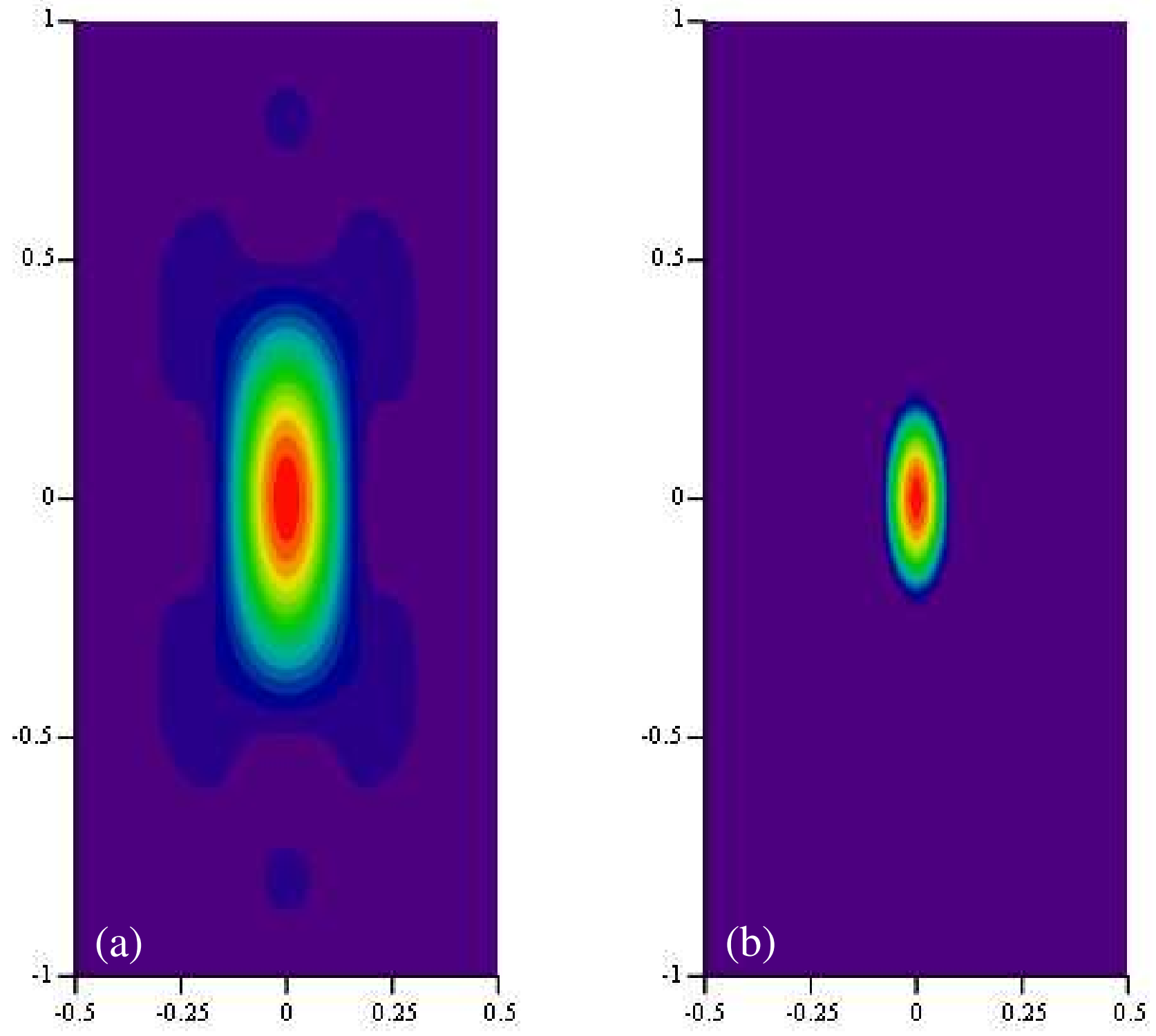